\documentclass[prc,aps,superscriptaddress,twocolumn,showpacs,nofootinbib]{revtex4-1}

\usepackage{graphicx,amsmath,amssymb,bm,multirow}
\usepackage{amsthm,mathrsfs}
\usepackage{amsfonts}
\usepackage{dcolumn}
\usepackage{bm}
\usepackage[usenames,dvipsnames]{color}
\graphicspath{{figures/}}

\newcommand{\beq}{\begin{equation}}
\newcommand{\eeq}{\end{equation}}
\newcommand{\beqn}{\begin{eqnarray}}
\newcommand{\eeqn}{\end{eqnarray}}

\newcommand{\bsub}{ \begin{subequations}}
\newcommand{\esub}{ \end{subequations}}

\renewcommand{\vec}[1]{\mbox{\boldmath $#1$}}

\begin{document}
\title{Transition from vibrational to rotational characters in low-lying
states of hypernuclei}
\author{H. Mei}
\affiliation{Department of Physics, Tohoku University, Sendai 980-8578, Japan}
\affiliation{Department of Physics and Astronomy, University of North Carolina, Chape Hill 27599-3255, USA}
\author{K. Hagino}
\affiliation{Department of Physics, Tohoku University, Sendai 980-8578, Japan}
\affiliation{Research Center for Electron Photon Science, Tohoku University, 1-2-1 Mikamine, Sendai 982-0826, Japan}
\affiliation{
National Astronomical Observatory of Japan, 2-21-1 Osawa,
Mitaka, Tokyo 181-8588, Japan}

\author{J. M. Yao}
\affiliation{Department of Physics and Astronomy, University of North Carolina, Chape Hill 27599-3255, USA}
\affiliation{School of Physical Science and Technology,
             Southwest University, Chongqing 400715, China}
\author{T. Motoba}
\affiliation{Laboratory of Physics, Osaka Electro-Communications University,
             Neyagawa 572-8530, Japan}
\affiliation{Yukawa Institute for Theoretical Physics, Kyoto University, Kyoto 606-8502, Japan }

\begin{abstract}
In order to clarify the nature of hypernuclear low-lying states,
we carry out a comprehensive study for the structure of
$^{144-154}_{~~~~~~~~\Lambda}$Sm-hypernuclei, which exhibit a transition
from vibrational to
rotational characters as the neutron number increases.
To this end, we employ
a microscopic particle-core coupling scheme based on a
covariant density functional theory.
We find that the positive-parity ground-state band in the hypernuclei shares a
similar structure to that of the corresponding
core nucleus.
That is,
regardless of whether the core nucleus is spherical or deformed,
each hypernuclear state is dominated
by the single configuration of the $\Lambda$ particle in the $s_{1/2}$ state
($\Lambda s_{1/2}$) coupled to one core state of the ground band.
In contrast, the low-lying negative-parity states
mainly consist of $\Lambda p_{1/2}$ and
$\Lambda p_{3/2}$ configurations coupled to plural nuclear core states.
We show that, while
the mixing amplitude between these configurations
is negligibly small in spherical and
weakly-deformed nuclei,
it strongly increases
as the core nucleus undergoes a transition to a well-deformed shape, being consistent
with the Nilsson wave functions.
We demonstrate that
the structure of these negative-parity states with spin $I$ can be well understood
based on the $LS$ coupling scheme, with the total orbital angular momentum of
$L=[I\otimes 1]$ and the spin angular momentum of $S=1/2$.
\end{abstract}

\maketitle

\section{Introduction}

With the advent of the high-resolution Ge detector array, ``Hyperball",
the  $\gamma$-ray spectroscopy has been carried out
for many hypernuclei \cite{HT06}. We particularly mention the measurement
for $^{13}_{~\Lambda}$C, which has provided a quantitative information
on the spin-orbit splittings~\cite{Ajimura01}.
In this experiment, the energy difference between the $1/2^-$ and
$3/2^-$ states was determined to be $152\pm54 \pm36$~keV,
which has been interpreted as the spin-orbit splitting between
1$p_{1/2}$ and 1$p_{3/2}$ hyperon states in $^{13}_{~\Lambda}$C.
This measurement, together with other measurements, thus has provided
a solid evidence for that the spin-orbit splitting of hyperon states
is smaller than that of nucleons,
by more than one order of magnitude~\cite{HT06},
which had been explained theoretically in terms of several
different mechanisms~\cite{BW77,Noble80,BB81,Boussy81}.

A similar interpretation in other hypernuclei may need a caution, however.
That is, the previous studies~\cite{Motoba85,Mei2014,Mei2015,Mei2016}
have demonstrated that
most $\Lambda$ hypernuclei are different from the $^{13}_{~\Lambda}$C,
in a sense that the lowest $1/2^-$ and $3/2^-$ states cannot
be naively interpreted as pure 1$p_{1/2}$ and 1$p_{3/2}$ hyperon states,
respectively, due to a large effect of configuration mixings.
This perturbs an interpretation of
their energy difference as the spin-orbit splitting for the $p$ orbits.
It has been shown that the amplitudes of the configuration mixing
depend much on the the collective properties of the core
nuclei \cite{Motoba83,Bando83, Hiyama00}.

In this paper, we investigate systematically
the nature of configuration mixing in several hypernuclei which
exhibit different collective properties.
The Sm isotopes around $N\sim 90$ provide an ideal
playground for this purpose,
even though the production of these hypernuclei may still be difficult
at this moment,
since it is well known that they exhibit
a shape phase transition from vibrational to rotational
characters as the number of neutron increases~\cite{Casten01}.
To this end, we shall use the microscopic particle-core coupling scheme,
based on the covariant density functional theory.
It is worth mentioning that
a covariant density functional theory has been successfully
applied to the shape phase transition in ordinary Sm nuclei
~\cite{Meng05,Niksic07,Li09N90}.

The paper is organized as follows. In Sec.~\ref{Sec:framework},
we briefly introduce the microscopic particle-core scheme for hypernuclei,
which uses results of the multi-reference covariant density
functional theory
for nuclear core excitations.
In Sec.~\ref{Sec:Results},
we apply this method to
the low-lying states of the Sm isotopes as well as the Sm $\Lambda$
hypernuclei, and discuss the nature of low-lying collective states in
these hypernuclei. We particularly discuss how the configuration mixing alters
as the shape of a core nucleus changes from spherical to deformed.
We then summarize the paper in Sec.~\ref{Sec:Summary}.

\section{method}
\label{Sec:framework}

\subsection{Multi-reference covariant density functional theory for nuclear core excitations}

We describe low-lying states of hypernuclei in the particle-core coupling
scheme. The first step in this method is to construct
the low-lying states of the core nuclei.
To this end, we adopt the multi-reference covariant density
functional theory (MR-CDFT)~\cite{Yao10,Yao14}.
In the MR-CDFT, the wave function of each nuclear core state is obtained
as a superposition of a set of quantum-number projected
mean-field reference states, $\vert \beta\rangle$.
Here, the reference states $\vert \beta\rangle$
are obtained with
deformation constrained relativistic mean-field plus BCS calculations with
the quadrupole deformation parameter $\beta$.
These states are then projected onto states with a good quantum number of
angular momentum and particle number as,
\begin{equation}
\label{projwf}
 \vert\Phi_{IM}(\beta)\rangle=\hat{P}^I_{MK}\hat{P}^N \hat{P}^Z
\vert \beta \rangle,
\end{equation}
where $\hat{P}^I_{MK}$ is the angular momentum projection operator, and
$\hat{P}^N$ and $\hat{P}^Z$ are the particle number projection operators
for neutron and proton, respectively.
For simplicity, we impose axial symmetry on the reference mean-field
states, and thus the $K$ quantum number in Eq. (\ref{projwf}) is
zero. Following the philosophy of the generator coordinate method (GCM),
the wave functions in the MR-CDFT are then constructed by superposing
the projected wave functions as,
\begin{equation}
  \label{GCMwf}
 \vert\Phi_{nIM}\rangle=\sum_{\beta}  F_{nI}(\beta)
 \vert\Phi_{IM}(\beta)\rangle.
 \end{equation}
Here, the weight function $F_{nI}(\beta)$ and the energy for
the state $\vert\Phi_{nI}\rangle$ are obtained by solving
the Hill-Wheeler-Griffin (HWG) equation, which is
derived from the variational principle~\cite{Ring80}.

In this paper,
we calculate the Hamiltonian kernel in the HWG equation
with the mixed-density prescription.
That is, we assume
the same functional form
for the off-diagonal elements of the energy overlap
(sandwiched by two different reference states) as that for the
diagonal elements, by replacing
all the densities and currents with the mixed ones ~\cite{Yao09,Yao10}.
In the calculations shown below, we adopt
the PC-F1 parametrization~\cite{BurRMFPC02} for the
relativistic point-coupling energy functional.

\subsection{Microscopic particle-core coupling scheme for $\Lambda$ hypernuclei}

We next construct the hypernuclear wave functions by expanding them
on the nuclear core states, Eq. (\ref{GCMwf}), which provide a set of
basis. The resultant wave functions read,
\begin{equation}
 \label{wavefunction}
 \displaystyle \Psi_{JM}(\vec{r},\{\vec{r}_i\})
 =\sum_{n,\ell, j, I}  {\mathscr R}_{j\ell n I}(r) [{\mathscr Y}_{j\ell }(\hat{\vec{r}})\otimes
\Phi_{n I}(\{\vec{r}_i\})]^{(JM)},
\end{equation}
where $\vec{r}$ and $\vec{r}_i$ are the coordinates of the $\Lambda$
hyperon and the
nucleons, respectively.
${\mathscr R}_{j\ell n I}(r)$ and
${\mathscr Y}_{j\ell}(\hat{\vec{r}})$ are the radial wave
function and the spin-angular wave function for the $\Lambda$-particle,
respectively.
The index $n = 1, 2, \ldots$ distinguishes different core states
with the same angular momentum $I$. In our previous publications
\cite{Mei2014,Mei2015,Mei2016}, we
have called this method ``the microscopic particle-rotor model'',
since we have mainly
considered couplings of a $\Lambda$ particle to rotational states of
deformed core nuclei. We instead call it ``the microscopic
particle-core coupling
scheme'' in this paper, as we deal with both spherical and deformed core
nuclei.

The radial wave functions, ${\mathscr R}_{j\ell n I}(r)$,
in Eq. (\ref{wavefunction})
and the energy of the
state, $E_J$, are obtained by solving the equation
$\hat{H}|\Psi_{JM}\rangle=E_J|\Psi_{JM}\rangle$.
We assume that the Hamiltonian $\hat H$ for the whole $\Lambda$ hypernucleus
is given by~\cite{Mei2016},
\begin{equation}
\hat H =\hat T_\Lambda +  \hat H_{\rm c}+ \sum^{A_c}_{i=1} \hat{V}^{N\Lambda}
(\vec{r},\vec{r}_{i}),
\label{eq:H}
\end{equation}
where $\hat T_\Lambda$ is the relativistic kinetic energy for the $\Lambda$ particle and
$\hat H_c$ is the many-body Hamiltonian for the core nucleus,
satisfying $\hat H_c \vert\Phi_{n I}\rangle= E_{n I} \vert\Phi_{n I}\rangle$.
The last term on the right side of Eq. (\ref{eq:H}) represents
the interaction term between the
$\Lambda$ particle and the nucleons in the core nucleus,
where $A_c$ is the mass number
of the core nucleus.
We here use
the $N\Lambda$ interaction derived from the point-coupling energy
functional with the PCY-S4 parametrization~\cite{Tanimura2012},
\begin{align}
\label{Scalar}
\hat{V}^{N\Lambda}_{\rm S}(\vec{r},\vec{r}_i)=& \alpha_S^{N\Lambda} \gamma^0_\Lambda
\delta(\vec{r}-\vec{r}_i)\gamma^0_N
+\delta_S^{N\Lambda}\gamma^0_\Lambda
\Big[\overleftarrow{\nabla}^2 \delta(\vec{r}-\vec{r}_i) \nonumber \\
+& \delta(\vec{r}-\vec{r}_i)\overrightarrow{\nabla}^2+ 2 \overleftarrow{\nabla}
\cdot\delta(\vec{r}-\vec{r}_i)
 \overrightarrow{\nabla}\Big]\gamma^0_N,\\
 \label{Vector}
\hat{V}^{N\Lambda}_{\rm V}(\vec{r},\vec{r}_i)=&\alpha_V^{N\Lambda} \delta(\vec{r}-\vec{r}_i)
+\delta_V^{N\Lambda} \Big[\overleftarrow{\nabla}^2 \delta(\vec{r}-\vec{r}_i)
 \nonumber \\
+& \delta(\vec{r}-\vec{r}_i) \overrightarrow{\nabla}^2+ 2
\overleftarrow{\nabla}\cdot\delta(\vec{r}-\vec{r}_i) \overrightarrow{\nabla}\Big],\\
 \label{Tensor}
\hat{V}^{N\Lambda}_{\rm Ten}(\vec{r},\vec{r}_i)=&
i\alpha_T^{N\Lambda}\gamma^0_\Lambda\Big[\overleftarrow{\nabla} \delta(\vec{r}-\vec{r}_i)
+\delta(\vec{r}-\vec{r}_i)\overrightarrow{\nabla}\Big]\cdot \vec{\alpha}.
\end{align}

In practice, the equation $\hat{H}|\Psi_{JM}\rangle=E_J|\Psi_{JM}\rangle$
is transformed into coupled-channels equations in the
relativistic framework,
in which all the diagonal and off-diagonal potentials
are determined from the MR-CDFT calculation.
We solve the coupled-channels equations by expanding
the four-component radial wave function ${\mathscr R}_{j\ell n I}(r)$
on a spherical harmonic oscillator basis.
See Refs.~\cite{Mei2014,Mei2015,Mei2016} for more details on
the framework.

\section{Results and discussion}
\label{Sec:Results}

Let us now apply the microscopic particle-core coupling
scheme to Sm hypernuclei and
discuss their low-lying collective states.
To this end, we generate
the reference states $\vert \beta\rangle$
by expanding the single-particle wave functions on a
harmonic oscillator basis with 12 major shells.
In the particle-number and angular-momentum projection calculations,
we choose the number of mesh points to be 9 for the gauge angle in $[0,\pi]$, and
16 for the Euler angle $\theta$ in the interval $[0,\pi]$.
In the coupled-channels calculations,
we include up to $l_{\rm max}=19$, $j_{\rm max}=37/2$, and $n_{\rm max}=3$ in the total wave function, Eq. (\ref{wavefunction}).
For each $j, l, n,$ and $I$, we include up to 18 (19) major
shells in the expansion of the upper (lower) component
of radial wave functions on the spherical harmonic
oscillator basis.

\subsection{Shape transition in Sm isotopes}
\label{Sm}

\begin{figure}[]
  \centering
  \includegraphics[width=8.5cm]{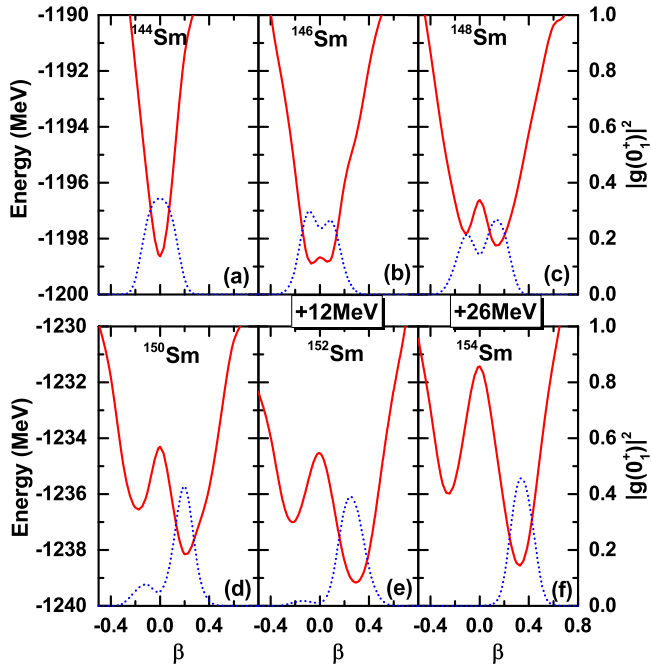}
\caption{The total energy in the mean-field approximation for Sm
isotopes as a function of the the intrinsic
quadrupole deformation $\beta$ (the solid lines).
The energy curves for $^{146}$Sm and $^{152}$Sm
are shifted upward by 12 MeV, while those for
$^{148}$Sm and $^{154}$Sm are shifted by 26 MeV.
The square of the collective wave functions, $|g_{nI}(\beta)|^2$, for
the GCM ground states ($0_1^+$) are also shown with the dashed curves. }
\label{EPSSm}
\end{figure}

Before we discuss the structure of the hypernuclei, we first discuss the
structure of the core nuclei.
The solid lines in Fig. \ref{EPSSm} show the total mean-field energy for the
$^{144-154}$Sm nuclei as a function of the quadrupole deformation
parameter, $\beta$.
For the $^{144}$Sm nucleus, one can see that the potential energy curve
is almost parabolic centered at the spherical shape.
As the neutron number increases,  the energy curve gradually
presents two pronounced minima, located on the oblate and the prolate sides,
respectively.
A previous study~\cite{Li09N90} has shown that these
two minima are connected with a tunneling along the trixiality (that is,
a $\gamma$ deformation).
In other words, the oblate minimum is actually a saddle point on
the energy surface. The true energy minimum is thus the prolate one,
that shifts gradually towards a large $\beta$ as the neutron
number increases, from $\beta=0.08$ for $^{148}$Sm to $\beta=0.32$ for $^{154}$Sm.

The square of the collective wave function, $|g_{nI}(\beta)|^2$, for
the ground state ($0^+_1$) of each isotope is shown by the dashed lines
in the figure.
Here, the collective wave function is defined as~\cite{Ring80},
\begin{equation}
\label{collectiveWF}
 g_{nI}(\beta) \equiv\sum_{\beta'} \big[ \mathcal{N}^{I}\big]^{1/2}(\beta,\beta')F_{nI}(\beta'),
\end{equation}
where the norm kernel is given
as $\mathcal{N}^{I}(\beta,\beta')\equiv\langle \beta\vert
\hat P^I\hat P^N\hat P^Z\vert \beta'\rangle$.
Notice that the weight function $F_{nI}(\beta)$ in Eq. (\ref{GCMwf})
cannot be interpreted as a probability amplitude due to the
non-orthgonality of the reference wave functions.
The figure clearly indicates that
the predominant component in the ground state changes
gradually from the spherical configuration
in $^{144}$Sm to the prolate deformed one in $^{154}$Sm.

\begin{figure}[]
  \centering
  \includegraphics[width=9cm]{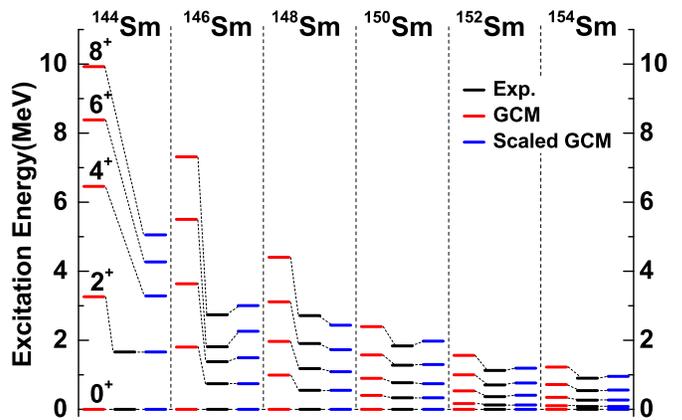}
\caption{The yrast levels of the Sm isotopes
calculated with the MR-CDFT (the red lines)
in comparison to the experiment data (the black lines)
taken from Ref.~\cite{NNDC}.
The figure also shows
the scaled levels (the blue lines) with a multiplicative factor of
$f=E^{\mathrm{exp}.}_{2^+}/E^{\mathrm{MR-CDFT}}_{2^+}$, that is
$E^{\mathrm{Scaled}}_{I^+}=f\cdot E^{\mathrm{MR-CDFT}}_{I^+}$. }
\label{SmSpetrum}
\end{figure}

Figure~\ref{SmSpetrum} shows calculated energy spectra for the
lowest $0^+$, $2^+$, $4^+$, $6^+$, and $8^+$ states in the Sm isotopes,
in comparison to the
corresponding data.
As one can see,
the main characters of the energy spectra are reasonably reproduced in this
calculation, although the excitation energies are somewhat overestimated.
In fact,
the energy spectra become close to the data, as shown in the figure,
after all the excitation energies are scaled so that the
experimental energy is reproduced for the first $2^+$ state.

\begin{figure}[]
  \centering
  \includegraphics[width=8.5cm]{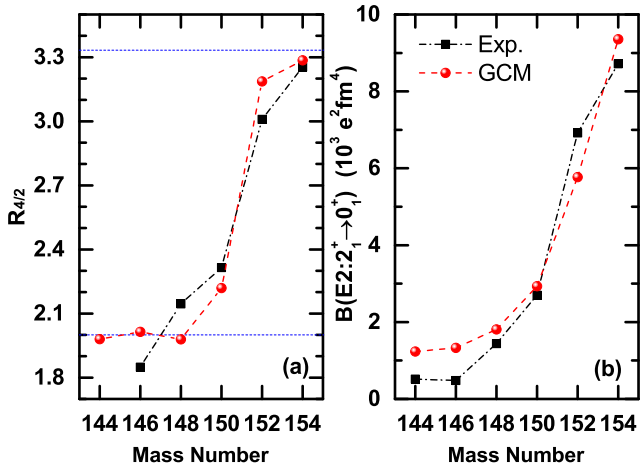}
\caption{(a) The ratio $R_{4/2}$
of the excitation energy for the first 4$^+$ state
to that for the first 2$^+$ state for the Sm isotopes
as a function of the mass number.
(b) The electromagnetic transition strength, in units
$10^3\,e^2$fm$^4$,
from the first 2$^+$
state to the ground state, $B(E2: 2^+_1\rightarrow 0^+_1)$,
as a function of the mass number for the Sm isotopes.
The experiment data are taken from Ref. \cite{NNDC}.}
  \label{E2BE2DV}
\end{figure}

In literature,
the energy ratio, $R_{4/2}\equiv E(4^+_1)/E(2^+_1)$,
of the excitation energy for the $4^+_1$ state to that for the $2^+_1$ state
has often been adopted to characterize nuclear collective excitations.
This value for $^{144}$Sm, $^{146}$Sm and $^{148}$Sm is
$R_{4/2}=1.98$, $2.01$ and $1.98$,
respectively, all of which are close to the value in the
harmonic oscillator limit, $R_{4/2}=2.0$.
With the increase of the neutron number, the value of $R_{4/2}$ increases
up to 3.29 for $^{154}$Sm, that is close to the value in the rigid rotor
limit, $R_{4/2}=3.33$.
Figure \ref{E2BE2DV}(a) illustrates this transition.
The figure clearly demonstrates that the theoretical results present
shape transition from spherical to deformed in the Sm isotopes around
$N=90$, which is in good agreement with the experimental
data.
This picture is verified also from the mass number dependence
of the electric quadrupole transition strength from the first 2$^+$ state
to the ground state, $B(E2; 2^+_1\rightarrow 0^+_1)$, as shown in
Fig.~\ref{E2BE2DV}(b).

\subsection{Low-lying spectrum of $_\Lambda$Sm isotopes}
\label{sec:SmLambda}

 \begin{figure}[]
  \includegraphics[width=9cm]{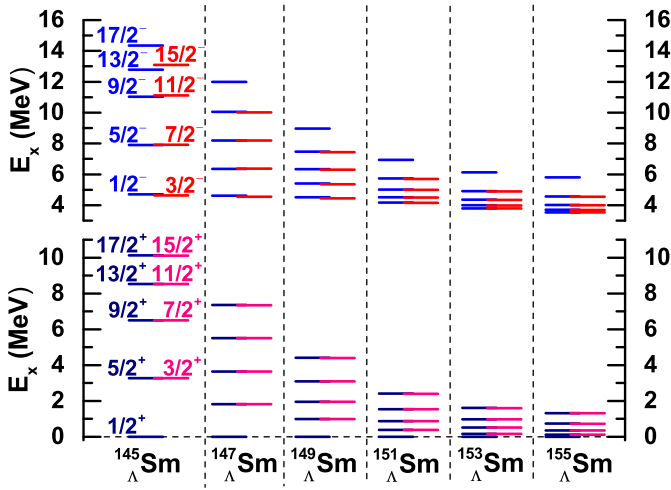}
  \vspace{-0.5cm}\caption{The spectrum for the first positive parity
states in the $_\Lambda$Sm isotopes.
The location of the spin doublet states are arranged
based on the yrast levels of the even Sm core nuclei.}
  \label{SmLSpectrum1}
\end{figure}

\begin{table}[htb]
\centering
\tabcolsep=3pt
 \caption{The probability $P$ of the dominant components, defined as  $P\equiv\int dr r^2 \vert {\mathscr R}_{j\ell n I}(r)\vert^2$, in the wave functions for the positive-parity states.}
  \begin{tabular}{cc|cccccc}
  \hline\hline
$J^\pi$    &$(lj)\otimes I^\pi_{n}$&$^{145}_{~~\Lambda}$Sm &$^{147}_{~~\Lambda}$Sm &$^{149}_{~~\Lambda}$Sm &$^{151}_{~~\Lambda}$Sm &$^{153}_{~~\Lambda}$Sm & $^{155}_{~~\Lambda}$Sm \\
     \hline
$1/2^+_1 $  &$s_{1/2}\otimes0_1^+$ &$0.998$ &$0.998$ &$0.997$ &$0.994$ &$0.988$& $0.982$ \\
$3/2^+_1 $  &$s_{1/2}\otimes2_1^+$ &$0.997$ &$0.997$ &$0.996$ &$0.993$ &$0.988$& $0.982$ \\
$5/2^+_1 $  &$s_{1/2}\otimes2_1^+$ &$0.997$ &$0.997$ &$0.996$ &$0.993$ &$0.988$& $0.982$ \\
$7/2^+_1 $  &$s_{1/2}\otimes4_1^+$ &$0.991$ &$0.996$ &$0.996$ &$0.993$ &$0.987$& $0.982$ \\
$1/2^+_2 $  &$s_{1/2}\otimes0_2^+$ &$0.987$ &$0.983$ &$0.993$ &$0.992$ &$0.987$& $0.990$ \\
$3/2^+_2 $  &$s_{1/2}\otimes2_2^+$ &$0.981$ &$0.996$ &$0.995$ &$0.992$ &$0.986$& $0.989$ \\
$5/2^+_2 $  &$s_{1/2}\otimes2_2^+$ &$0.980$ &$0.996$ &$0.995$ &$0.991$ &$0.986$& $0.989$\\
$7/2^+_2 $  &$s_{1/2}\otimes4_2^+$ &$0.988$ &$0.993$ &$0.994$ &$0.987$ &$0.985$& $0.986$\\
\hline\hline
 \end{tabular}
    \label{component_SmL1}
   \end{table}

We now discuss the low-lying states in the $_{\Lambda}$Sm hypernuclei, in which a
$\Lambda$ particle couples to the core states presented in the previous
subsection.
Figure~\ref{SmLSpectrum1} shows the calculated yrast positive-parity
states in the $_\Lambda$Sm isotopes.
The probability for the dominant configuration in the wave function
for the $1/2^+$, $3/2^+$, $5/2^+$, and $7/2^+$ states is also
presented in Table \ref{component_SmL1}.
These positive-parity states
are dominated by the configuration of
$[\Lambda_{s_{1/2}}\otimes I^+_1]$ with the weight around $99\%$,
where $\Lambda_{s_{1/2}}$ denotes the $\Lambda$ particle in the $s_{1/2}$
configuration.
They are nontrivial results obtained for all the $_{\Lambda}$Sm isotopes.
These states
have a similar excitation energy to that of the nuclear core state
with $I^+_1$, and are nearly two-fold degenerate except for the $1/2^+_1$
state.
These characters are similar to hypernuclei in the
light-mass region, and are also consistent with our
previous calculation for $^{155}_{~~\Lambda}$Sm with a simplified
$\Lambda N$ interaction~\cite{Mei2015}.

 \begin{figure*}[htb]
  \includegraphics[width=17cm]{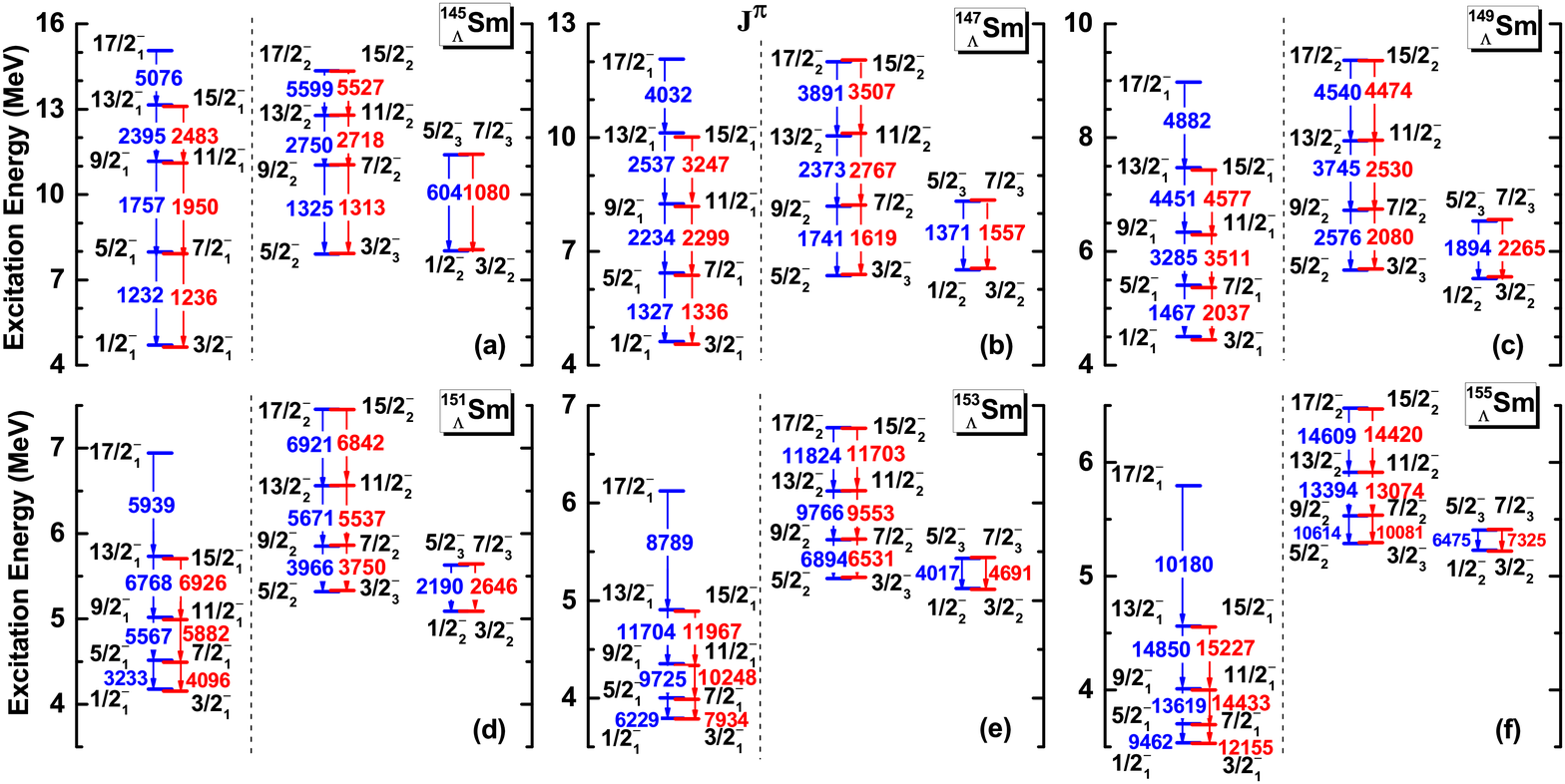}
  \vspace{-0.5cm}\caption{The low-lying negative-parity states in the
$_\Lambda$Sm isotopes obtained with the microscopic particle-core
coupling scheme based on the covariant density functional theory.
The arrows indicate the E2 transition strengths, given in units
of $e^2$ fm$^4$.}
  \label{SmLSpectrum2}
\end{figure*}

\begin{table}[htb]
\centering
\tabcolsep=2pt
 \caption{Same as Table \ref{component_SmL1},
but for the negative-parity states shown in Fig. \ref{SmLSpectrum2}.
The blank entries indicate the probabilities smaller than $0.001$.}
\begin{tabular}{cc|cccccc}
  \hline\hline
$J^\pi$    &$(lj)\otimes I^\pi_{n}$&$^{145}_{~~\Lambda}$Sm &$^{147}_{~~\Lambda}$Sm &$^{149}_{~~\Lambda}$Sm &$^{151}_{~~\Lambda}$Sm &$^{153}_{~~\Lambda}$Sm & $^{155}_{~~\Lambda}$Sm
\\     \hline
$1/2^-_1 $  &$p_{1/2}\otimes0_1^+$ &$0.986$ &$0.964$ &$0.859$ &$0.484$ &$0.348$& $0.322$ \\
  $$        &$p_{3/2}\otimes2_1^+$ &$0.012$ &$0.033$ &$0.136$ &$0.503$ &$0.627$& $0.639$ \\

$3/2^-_1 $  &$p_{1/2}\otimes2_1^+$ &$0.006$ &$0.015$ &$0.054$ &$0.204$ &$0.271$& $0.281$     \\
            &$p_{3/2}\otimes0_1^+$ &$0.986$ &$0.965$ &$0.876$ &$0.545$ &$0.395$& $0.363$ \\
            &$p_{3/2}\otimes2_1^+$ &$0.006$ &$0.017$ &$0.064$ &$0.238$ &$0.309$& $0.318$    \\

$5/2^-_1 $  &$p_{1/2}\otimes2_1^+$ &$0.980$ &$0.959$ &$0.573$ &$0.453$ &$0.385$& $0.346$          \\
            &$p_{3/2}\otimes4_1^+$ &$0.012$ &$0.034$ &$0.154$ &$0.377$ &$0.462$& $0.504$ \\
            &$p_{3/2}\otimes2_1^+$ &$$      &$$      &$0.262$ &$0.156$ &$0.127$& $0.112$     \\

$7/2^-_1 $  &$p_{1/2}\otimes4_1^+$ &$0.008$ &$0.022$ &$0.074$ &$0.183$ &$0.232$& $0.258$         \\
            &$p_{3/2}\otimes2_1^+$ &$0.980$ &$0.954$ &$0.854$ &$0.653$ &$0.554$& $0.497$ \\
            &$p_{3/2}\otimes4_1^+$ &$0.006$ &$0.018$ &$0.062$ &$0.150$ &$0.188$& $0.207$    \\

 $9/2^-_1 $ &$p_{1/2}\otimes4_1^+$ &$0.843$ &$0.931$ &$0.570$ &$0.481$ &$0.398$& $0.372$ \\
            &$p_{3/2}\otimes4_1^+$ &$0.071$ &$0.016$ &$0.288$ &$0.210$ &$0.166$& $0.154$ \\
            &$p_{3/2}\otimes6_1^+$ &$0.040$ &$0.033$ &$0.131$ &$0.295$ &$0.407$& $0.437$ \\ \hline

 $1/2^-_2 $ &$p_{1/2}\otimes0_1^+$ &$0.013$ &$0.035$ &$0.133$ &$0.498$ &$0.635$& $0.654$\\
            &$p_{3/2}\otimes2_1^+$ &$0.978$ &$0.945$ &$0.813$ &$0.472$ &$0.351$& $0.330$\\

$3/2^-_2 $   &$p_{1/2}\otimes2_1^+$ &$0.734$ &$0.655$  &$0.272$  &$0.210$ &$0.136$& $0.119$ \\
             &$p_{3/2}\otimes2_1^+$ &$0.246$ &$0.295$  &$0.583$  &$0.322$ &$0.266$& $0.258$ \\
             &$p_{3/2}\otimes0_1^+$ &$0.012$ &$0.031$  &$0.113$  &$0.435$ &$0.583$& $0.605$\\

 $5/2^-_2 $ &$p_{1/2}\otimes2_1^+$ &$$      &$$      &$0.283$ &$0.221$ &$0.210$& $0.202$\\
            &$p_{3/2}\otimes2_1^+$ &$0.988$ &$0.978$ &$0.658$ &$0.746$ &$0.764$& $0.780$\\

 $7/2^-_2 $ &$p_{1/2}\otimes4_1^+$ &$0.261$ &$0.240$ &$0.580$ &$0.497$ &$0.478$& $0.484$\\
            &$p_{3/2}\otimes4_1^+$ &$0.695$ &$0.726$ &$0.205$ &$0.486$ &$0.492$& $0.499$\\
            &$p_{3/2}\otimes2_2^+$ &$0.014$ &$     $ &$0.139$ &$     $ &$     $& $     $\\

 $9/2^-_2 $ &$p_{1/2}\otimes4_1^+$ &$0.083$ &$0.013$ &$0.303$ &$0.281$ &$0.263$& $0.265$ \\
            &$p_{3/2}\otimes4_1^+$ &$0.888$ &$0.943$ &$0.619$ &$0.703$ &$0.707$& $0.719$ \\
             \hline

 $3/2^-_3 $ &$p_{1/2}\otimes2_1^+$ &$0.253$ &$0.313$ &$0.636$ &$0.553$ &$0.571$& $0.579$\\
            &$p_{3/2}\otimes2_1^+$ &$0.741$ &$0.672$ &$0.304$ &$0.416$ &$0.402$& $0.401$\\

 $5/2^-_3 $  &$p_{1/2}\otimes2_1^+$ &$0.007$ &$0.034$ &$0.123$   &$0.313$ &$0.389$& $0.429$ \\
             &$p_{3/2}\otimes4_1^+$ &$0.712$ &$0.759$ &$0.724$   &$0.609$  &$0.514$& $0.464$ \\

 $7/2^-_3 $  &$p_{1/2}\otimes4_1^+$ &$0.527$ &$0.476$ &$0.302$   &$0.307$ &$0.266$& $0.237$ \\
             &$p_{3/2}\otimes4_1^+$ &$0.251$ &$0.132$ &$0.368$   &$0.350$  &$0.294$& $0.272$ \\
             &$p_{3/2}\otimes2_2^+$ &$0.116$ &$0.269$ &$0.137$   &$$       &$$     & $$ \\
             &$p_{3/2}\otimes2_1^+$ &$0.010$ &$0.033$ &$0.124$   &$0.336$  &$0.428$& $0.472$ \\
\hline\hline
 \end{tabular}
    \label{component_SmL}
   \end{table}

In the negative parity states of $_{\Lambda}$Sm isotopes, novel and
interesting features are disclosed.
The low-lying negative-parity states are shown in Fig.~\ref{SmLSpectrum2}.
We summarize in Table~\ref{component_SmL}
the dominant components in the wave functions for a few selected levels.
One can see that
these negative-parity states are formed mainly from a $\Lambda$-particle
in $p$ orbitals coupled to core states, and are
nearly two-fold degenerate.
The E2 transition strengths between the negative parity states
are also shown in the figure.
The ratio of the $E2$ transition strength for the transition
$5/2_1^- \rightarrow 1/2_1^-$ to that for $9/2_1^- \rightarrow 5/2_1^-$ is
0.641 and 0.695 in $^{153}_{~~\Lambda}$Sm  and $^{155}_{~~\Lambda}$Sm, respectively.
These values are both close to 0.7, that is the ratio of the
$E2$ transition strength for $2^+ \rightarrow 0^+$ to that for
$4^+ \rightarrow 2^+$ in the $K=0$ ground-state rotational
band of well-deformed nuclei~\cite{Ring80}.
A simple relation among the $B(E2)$ values for the negative-parity bands
in a well deformed hypernucleus is further discussed in Appendix A.

\begin{figure}[]
  \centering
  \includegraphics[width=8.5cm]{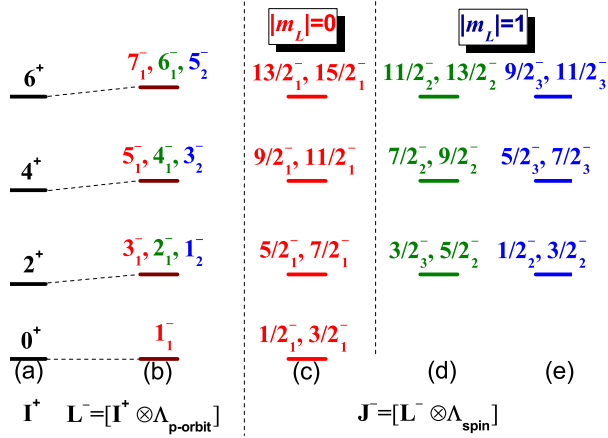}
\caption{A schematic picture for hypernuclear states
based on the $LS$ coupling scheme.
It is assumed that a
$\Lambda$-particle is in a $p$-orbital and
coupled to spherical core states.}
\label{schamtic}
\end{figure}

The level structure of the negative-parity states shown in
Fig. \ref{SmLSpectrum2}
can be understood in terms of the $LS$ coupling scheme.
To demonstrate this, let us consider a simplified situation in which
the core states with $K=0$ shown in Fig. ~\ref{schamtic} (a)
are coupled to a $\Lambda$ particle in $p$
orbitals. We first couple the core angular momentum $I$ with the
orbital angular momentum of the $\Lambda$ particle, $l_\Lambda=1$.
This results in the levels shown in Fig.~\ref{schamtic} (b). These
levels may be categorized according to the projection of the
total orbital angular momentum, $\vec{L}=\vec{I}+\vec{l}_\Lambda$,
on to the symmetric axis, that is,
$m_L=K+m_\Lambda$. Since $K$=0, there are two possibilities for $|m_L|$,
that is, $m_L = 0$ and $\pm 1$. The levels with $m_L=0$ form a band with
$1_1^-, 3_1^-, 5_1^-, 7_1^-, \cdots$,
while the levels with $m_L=\pm 1$
form another band with
$1_2^-, 2_1^-, 3_2^-, 4_1^-, \cdots$.
If one further couples the spin 1/2 of the $\Lambda$ particle to these
rotational bands, one obtains the levels in Fig.~\ref{schamtic} (c)
for the $m_L=0$ band, and the levels in Figs.~\ref{schamtic} (d) and
(e) for the $m_L=\pm 1$ band.
For a given $I$, the levels belonging to
the $m_L=0$ and the $m_L=\pm 1$ bands are degenerate in energy
for spherical hypernuclei. For prolately
deformed hypernuclei, on the other hand, the level in the
$m_L=0$ band is lower in energy than the levels in the $m_L=\pm 1$ band
because the former configuration
gains more energy due to a better overlap with the core nucleus.
This feature explain well the energy relation among the
three doublet states of (5/2$_1^-$, 7/2$_1^-$), (3/2$_2^-$, 5/2$_2^-$),
and (1/2$_2^-$, 3/2$_2^-$) in, e.g., $^{145}_{~~\Lambda}$Sm and
$^{155}_{~~\Lambda}$Sm (see Fig.~\ref{SmLSpectrum2}).

\begin{figure}[]
  \centering
  \includegraphics[width=8.5cm]{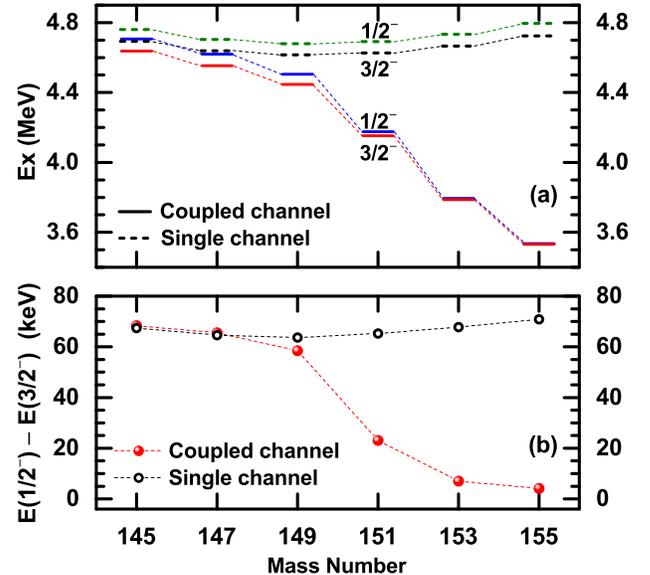}
\caption{(a) The energy levels of the $1/2^-_1$ and $3/2^-_1$ states
in the $_\Lambda$Sm hypernuclei as a function of the mass number.
The solid line and the dashed line indicate the results of
the coupled-channels and the single-channel calculations, respectively.
(b) The energy splitting between the $1/2^-_1$ and $3/2^-_1$ states
shown in the upper panel. }
  \label{SplitDevelop}
\end{figure}

Figure~\ref{SplitDevelop}(a)
shows in details the excitation energy
of the lowest $1/2^-_1$ and $3/2^-_1$ states
in the $_\Lambda$Sm isotopes
as a function of the neutron number.
The dashed energy levels show the results of the single-channel calculations, for
which the sum in Eq. (\ref{wavefunction}) is restricted only to a single
configuration.
For the lowest $1/2^-_1$ and $3/2^-_1$ states,
the configuration in the single-channel calculation is
a pure configuration of $[\Lambda p_{1/2} \otimes 0^+_1]$
and $[\Lambda p_{3/2} \otimes 0^+_1]$, respectively.
Their excitation energies are around 4.8 MeV for all the hypernuclei
considered in this paper, which is close to
the energy $\frac{2}{3}\times41A^{-1/3}\sim 5.14$ MeV with $A\sim 150$
for exciting one hyperon
from $s$ orbit to $p$ orbit.
The energy difference between these states remains around 70 keV, as shown
by the open circles in
Figure~\ref{SplitDevelop}(b).
In marked contrast,
the energy of the $1/2^-_1$ and $3/2^-_1$ states
obtained by including the configuration mixing effect
decreases
continuously from 4.7 MeV to 3.5 MeV as the neutron number
increases from 82 to 92 (see the solid lines in
Fig.~\ref{SplitDevelop}(a)).
The splitting of these two states also decreases from 68 keV to 4 keV,
as shown in Fig.~\ref{SplitDevelop}(b) by the filled circles.
The deviation from the single-channel calculations increases as
the core nucleus undergoes phase transition
from a spherical vibrator to a well-deformed rotor, indicating
a stronger configuration mixing effect in deformed hypernuclei.

 \begin{figure}[]
  \centering
  \includegraphics[width=8.5cm]{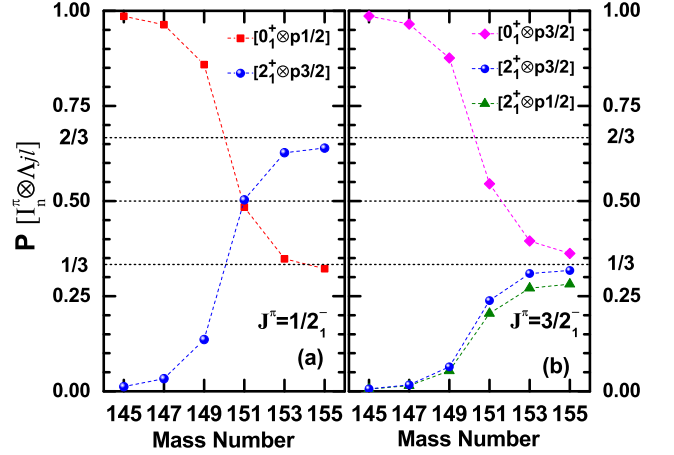}
\vspace{-0.1cm}\caption{The probability $P_k$ for the dominant
components in the wave function of (a) the $1/2^-_1$ state and (b) the
$3/2^-_1$ state as a function of the mass number of the
$_\Lambda$Sm isotopes.}
  \label{PDeV}
\end{figure}

This feature can be seen also in the
compositions of the wave functions listed in
Table~\ref{component_SmL}.
In $^{145}_{~~\Lambda}$Sm($^{147}_{\Lambda}$Sm), the $1/2^-_1$ and $3/2^-_1$ states are
almost pure configuration of $[\Lambda p_{1/2} \otimes 0^+_1]$
and $[\Lambda p_{3/2} \otimes 0^+_1]$, respectively.
This is consistent with the fact that the single-channel calculation
works well for this hypernucleus (see Fig.~\ref{SplitDevelop}).
With the increase of the neutron number, the mixing
between the $[\Lambda p_{1/2} \otimes 0^+_1]$
and $[\Lambda p_{3/2} \otimes 2^+_1]$ configurations
in the $1/2^-_1$ state
becomes stronger and reaches the largest value in $^{155}_{~~\Lambda}$Sm.
This feature is shown clearly in Fig.~\ref{PDeV}(a).
One can see that the mixing between $0^+_1$ and $2^+_1$ becomes
almost half-and-half in $^{151}_{~~\Lambda}$Sm.
In the well-deformed $^{155}_{~~\Lambda}$Sm, the weight for the
$[\Lambda p_{1/2}\otimes 0_1^+]$ configuration becomes  32.2\% while
that for the $[\Lambda p_{3/2}\otimes 2_1^+]$ configuration
becomes 63.9\%.
For the $3/2^-_1$ state, on the other hand,
the wave function shows a mixture of the
$[\Lambda p_{3/2}\otimes 0_1^+]$, $[\Lambda p_{1/2}\otimes 2_1^+]$
and $[\Lambda p_{3/2}\otimes 2_1^+]$ configurations.
The mass number dependence of the weight factors is shown
in Fig.~\ref{PDeV}(b), indicating a similar feature as in the $1/2_1^-$ state.
That is, the configuration mixing becomes stronger as the core nucleus
undergoes a transition from spherical to deformed.
For the $^{155}_{~~\Lambda}$Sm hypernucleus, the weight factors are
36.3\%,  28.1\%, and 31.8\%,
for the
$[\Lambda p_{3/2}\otimes 0_1^+]$, $[\Lambda p_{1/2}\otimes 2_1^+]$
and $[\Lambda p_{3/2}\otimes 2_1^+]$ configurations,
respectively.

It is worth mentioning that for all the negative-parity states
in the $m_L=0$ band of $^{155}_{~~\Lambda}$Sm shown in
Table~\ref{component_SmL}, the weight for the configurations
with $p_{1/2}$ is around 33\%, while a sum of the weight factors for the configurations
with $p_{3/2}$ is close to 67\%.
To understand this, let us employ the Nilsson model for the
hyperon with an axially deformed
potential,
$V(\vec{r})=V_0(r) - \beta R_0 \frac{d V_0(r)}{dr} Y_{20}(\hat{\vec{r}})$.
Notice that, with this deformed potential, several
orbital angular momenta $l$ and total angular momenta $j$
are mixed in the hyperon wave function.
Treating the deformed part of the potential,
$-\beta R_0 \frac{d V_0(r)}{dr} Y_{20}(\hat{\vec{r}})$,
with the first order perturbation theory and neglecting
the mixture across two major shells,
one can write the wave function for the lowest negative parity state as
$|\psi_{\Lambda}\rangle=C_{1}|p3/2,1/2\rangle+C_{2}|p1/2,1/2\rangle$, where
$|p3/2,1/2\rangle$ and $|p1/2,1/2\rangle$ are single-particle
wave functions in the spherical limit with $j_z=1/2$.
The coefficients $C_1$ and $C_2$ are simply determined by the
following eigenvalue equation,
\beqn
\label{eigenEq}
 &&\left(
  \begin{array}{cc}
    \langle \mathscr{Y}_{p3/2,1/2}|Y_{20}|\mathscr{Y}_{p3/2,1/2}\rangle & \langle \mathscr{Y}_{p3/2,1/2}|Y_{20}|\mathscr{Y}_{p1/2,1/2}\rangle \\
    \langle \mathscr{Y}_{p1/2,1/2}|Y_{20}|\mathscr{Y}_{p3/2,1/2}\rangle & \langle \mathscr{Y}_{p1/2,1/2}|Y_{20}|\mathscr{Y}_{p1/2,1/2}\rangle \\
  \end{array}
\right)\nonumber\\
&& \times\left(
         \begin{array}{c}
          C_1 \\
          C_2 \\
         \end{array}
       \right)
       =\lambda\left(
         \begin{array}{c}
          C_1 \\
          C_2 \\
         \end{array}
       \right).
\eeqn
The value of the matrix elements
$\displaystyle\langle \mathscr{Y}_{p3/2,1/2}|Y_{20}|\mathscr{Y}_{p3/2,1/2}\rangle$,
$\displaystyle\langle \mathscr{Y}_{p3/2,1/2}|Y_{20}|\mathscr{Y}_{p1/2,1/2}\rangle$,
$\displaystyle\langle \mathscr{Y}_{p1/2,1/2}|Y_{20}|\mathscr{Y}_{p3/2,1/2}\rangle$ and
$\displaystyle\langle \mathscr{Y}_{p1/2,1/2}|Y_{20}|\mathscr{Y}_{p1/2,1/2}\rangle$  is
$\displaystyle\frac{1}{\sqrt{20\pi}}$, $\displaystyle-\dfrac{2}{\sqrt{20\pi}}$,
$\displaystyle-\frac{2}{\sqrt{20\pi}}$ and 0, respectively.
The solutions of Eq.(\ref{eigenEq}) then give two eigenvectors,
$\displaystyle(C_1,C_2)^T = (\sqrt{2/3},-\sqrt{1/3})^T$ and
$\displaystyle(\sqrt{1/3},\sqrt{2/3})^T$,
with the eigenvalues of $\displaystyle\lambda= 2/\sqrt{20\pi}$
and $\displaystyle -1/\sqrt{20\pi}$, respectively.
For a positive value of $\beta$, the former state is lower in energy.
For this state, the probability of the
$p_{3/2}$ component reads 66.7\%  and that of $p_{1/2}$ component is 33.3\%.
This clearly implies that
the weight factors shown in Table~\ref{component_SmL} are consistent
with the Nilsson model and thus can be understood in terms of the
strong coupling limit of the particle-rotor model.

 \begin{figure}[]
  \centering
  \includegraphics[width=8cm]{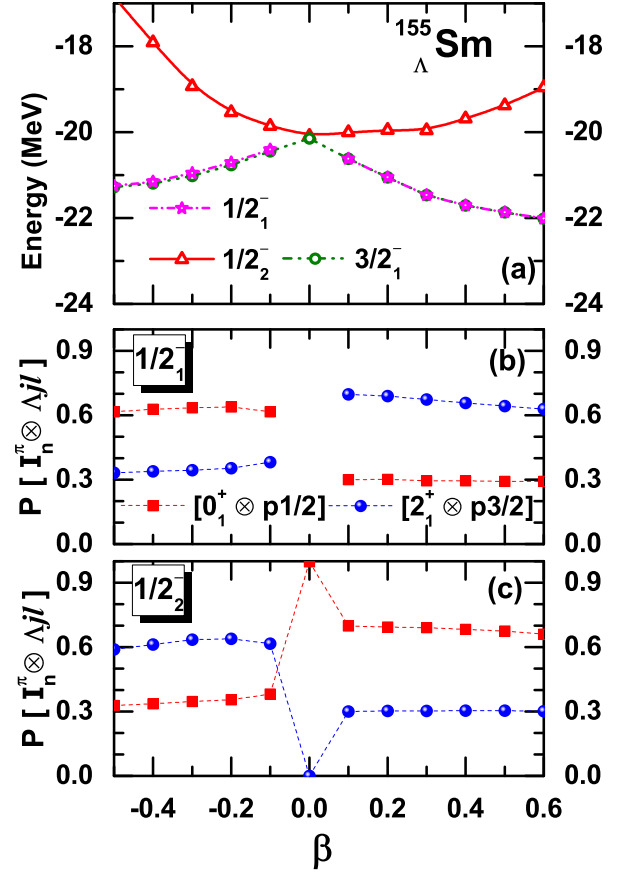}
\vspace{-0.1cm}\caption{(a) The energy of the $J^\pi=1/2^-_1,1/2^-_2$
and $3/2^-_1$ states of the $^{155}_{~~\Lambda}$Sm hypernucleus
as a function of the quadrupole deformation $\beta$ of the core nucleus.
These curves are obtained with the microscopic particle-rotor model
calculation by setting the excitation energies of the core states to be
zero. (b) The probability $P_k$ of the dominant components in the $1/2^-_1$
state as a function of the deformation parameter, $\beta$.
(c) Same as (b), but for the $1/2^-_2$ state. }
  \label{PESL}
\end{figure}

In addition, we also carry out the coupled-channels calculation for
$^{155}_{~~\Lambda}$Sm by setting the excitation energies of the
nuclear core states to be zero.
In order to draw the energy curve as a function of the deformation parameter,
we take $F_{nI}(\beta)=\delta_{\beta,\beta'}$ in Eq. (\ref{GCMwf}) and
compute the total energy for $\beta=\beta'$ \cite{Weixia15}.
Notice that this is a reasonable approximation for well-deformed hypernuclei.
The calculated energy for the $1/2^-_1, 1/2^-_2$ and $3/2^-_1$ states
are shown in Fig.~\ref{PESL}(a).
The splitting of the single-particle states due to nuclear
deformation is a well known feature of the Nilsson
diagram~\cite{Ring80}.
The main components of the wave function are shown
in Figs. ~\ref{PESL}(b) and (c) for the $1/2_1^-$ and
$1/2_2^-$ states, respectively.
One can see that the $1/2^-_1$ state composes mainly of the
$[\Lambda p_{1/2}\otimes 0_1^+]$ and $[\Lambda p_{3/2}\otimes 2_1^+]$
configurations with a rather constant mixing weight of
around 30\% and 70\%, respectively, on the prolate side.
The mixing weights for these two configurations are exchanged
on the oblate side. The mixing weights for the $1/2^-_2$ state are
just opposite to those for the $1/2^-_1$ state.
These findings confirm the analysis based on the simple Nilsson potential
presented in the previous paragraph.

\section{Summary}
\label{Sec:Summary}

We have systematically investigated the configuration mixing in
low-lying states of Sm hypernuclei
using the microscopic particle-core coupling scheme based on the
covariant density functional theory.
We emphasize that
this is the first microscopic calculation for hypernuclear
spectra in (medium-)heavy hypernuclei
and can be achieved only with the mean-field based calculations,
in which the beyond-mean-field correlations are also included.
We have found that the positive-parity ground-state band shares
a similar structure to that for the core nucleus.
That is, the hypernuclear states with spin-parity of $(I\pm1/2)^+$
are dominated by
the configuration of $[\Lambda s_{1/2}\otimes I^+]$, where
$\Lambda s_{1/2}$ denotes the
$\Lambda$ particle in the $s_{1/2}$ state, regardless of whether
the core nucleus is spherical or deformed.
In contrast, the low-lying negative-parity states show an admixture of the
$\Lambda p_{1/2}$ and the $\Lambda p_{3/2}$ configurations coupled with nuclear core
states having $I$ and $I\pm2$.
We have shown that the mixing amplitude is negligibly small
in spherical and weakly-deformed nuclei,
while it becomes increasingly stronger as the core nucleus undergoes
a shape transition to a well-deformed shape.
We have demonstrated that the energy spectra for low-lying negative
parity states in Sm hypernuclei
can be well understood with the $LS$ coupling scheme with the orbital
angular momentum of
$L=\vert I-1\vert, I, I+1$ and the spin angular momentum of $S=1/2$.
For well-deformed hypernuclei, the spectra as well as the wave
functions are also consistent with the Nilsson model.

The conclusion obtained in this paper can be applied to hypernuclei
in any mass region,
provided that the low-lying states of the core nucleus are dominated
by quadrupole collective excitations.
This indicates that the spin-orbit splitting for the hyperon
$p$-orbital should be estimated
from the energy difference between the first $1/2^-$ and $3/2^-$
states in hypernuclei with a nearly spherical nuclear core.

\section*{Acknowledgments}
This work was supported in part by the Tohoku University Focused Research
Project \lq\lq
Understanding the origins for matters in universe\rq\rq, JSPS KAKENHI
Grant Number 2640263. The National Natural Science Foundation of China under Grant Nos.
11575148, 11475140, 11305134.

\appendix

\section{$E2$ transition strengths in a well deformed hypernucleus}
\begin{figure}[htb]
  \centering
  \includegraphics[clip,width=8.5cm]{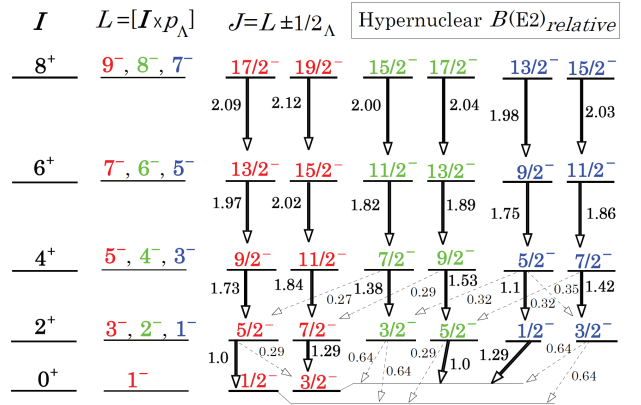}
\vspace{-0.1cm}
\caption{
The relation among the $E2$ transition strengths for transition
between low-lying negative parity states in a well deformed
hypernucleus. The solid line indicates strong transitions, for which
the $B(E2)$ value is larger than the $B(E2)$ value for the transition
from the first excited
state with $(I,L,J)=(2,3,5/2)$ to the ground state with
$(I,L,J)=(0,1,1/2)$, where $I$, $L$, and $J$ are the spin of the core
nucleus, the total orbital angular momentum of the hypernuclei and
the total spin of the hypernuclei, respectively.
The dashed lines denote weaker transitions.
See Table~\ref{table:B(E2)} for the actual values of the transition
strengths. We note that the band with $m_L=0$ should be lower in energy than the other two with $m_L=\pm 1$ for a well-deformed hypernucleus,
as we have discussed in Figs.~\ref{SmLSpectrum2} and \ref{schamtic}.}
\label{fig:B(E2)}
\end{figure}
\begin{table}[htb]
\centering
\tabcolsep=2pt
 \caption{
The $E2$ transition strengths for transition
between low-lying negative parity states in a well deformed
hypernucleus.
The $B(E2)$ values are given relative to
the value for the transition
from the first excited
state with $(I,L,J)=(2,3,5/2)$ to the ground state with
$(I,L,J)=(0,1,1/2)$, where $I$, $L$, and $J$ are the spin of the core
nucleus, the total orbital angular momentum of the hypernuclei and
the total spin of the hypernuclei, respectively.
Only the $B(E2)$ values larger than 0.25 are listed.
}
\begin{tabular}{ccc|ccc|c}
  \hline\hline
$I_i$ & $L_i$ & $J_i$ & $I_f$ & $L_f$ & $J_f$
& $B(E2:J_i\to J_f)$ \\
\hline
2 & 3 & 5/2 & 0 & 1 & 1/2 & 1.00 \\
2 & 3 & 5/2 & 0 & 1 & 3/2 & 0.286 \\
2 & 3 & 7/2 & 0 & 1 & 3/2 & 1.29 \\
2 & 2 & 3/2 & 0 & 1 & 1/2 & 0.643 \\
2 & 2 & 3/2 & 0 & 1 & 3/2 & 0.643 \\
2 & 2 & 5/2 & 0 & 1 & 1/2 & 0.286 \\
2 & 2 & 5/2 & 0 & 1 & 3/2 & 1.00 \\
2 & 1 & 1/2 & 0 & 1 & 3/2 & 1.29 \\
2 & 1 & 3/2 & 0 & 1 & 1/2 & 0.643 \\
2 & 1 & 3/2 & 0 & 1 & 3/2 & 0.643 \\
4 & 5 & 9/2 & 2 & 3 & 5/2 & 1.73 \\
4 & 5 & 11/2 & 2 & 3 & 7/2 & 1.84 \\
4 & 4 & 7/2 & 2 & 3 & 5/2 & 0.273 \\
4 & 4 & 9/2 & 2 & 3 & 7/2 & 0.289 \\
4 & 4 & 7/2 & 2 & 2 & 3/2 & 1.38 \\
4 & 4 & 9/2 & 2 & 2 & 5/2 & 1.53 \\
4 & 3 & 5/2 & 2 & 2 & 3/2 & 0.315 \\
4 & 3 & 7/2 & 2 & 2 & 5/2 & 0.354 \\
4 & 3 & 5/2 & 2 & 1 & 1/2 & 1.10 \\
4 & 3 & 5/2 & 2 & 1 & 3/2 & 0.315 \\
4 & 3 & 7/2 & 2 & 1 & 3/2 & 1.42 \\
6 & 7 & 13/2 & 4 & 5 & 9/2 & 1.97 \\
6 & 7 & 15/2 & 4 & 5 & 11/2 & 2.02 \\
6 & 6 & 11/2 & 4 & 4 & 7/2 & 1.82 \\
6 & 6 & 13/2 & 4 & 4 & 9/2 & 1.89 \\
6 & 5 & 9/2 & 4 & 3 & 5/2 & 1.75 \\
6 & 5 & 11/2 & 4 & 3 & 7/2 & 1.86 \\
\hline\hline
 \end{tabular}
\label{table:B(E2)}
 \end{table}

In this Appendix, we derive a simple expression for the $B(E2)$ values
for the $E2$ transition between the negative-parity states in a well
deformed hypernucleus. To this end, we use the $LS$ coupling scheme
discussed in Sec.~\ref{sec:SmLambda} and consider a transition from
the initial state $|J_iM_i\rangle$ to the final state $|J_fM_f\rangle$,
whose wave function is given by
\begin{equation}
|J_iM_i\rangle = \left|\left[[I_i\otimes l_\Lambda]^{(L_i)}\otimes s_\Lambda
\right]^{(J_iM_i)}\right\rangle,
\end{equation}
\begin{equation}
|J_fM_f\rangle = \left|\left[[I_f\otimes l_\Lambda]^{(L_f)}\otimes s_\Lambda
\right]^{(J_fM_f)}\right\rangle,
\end{equation}
respectively.
Here, $I_i$ and $I_f$ are the initial and the final spin of the core nucleus,
respectively, and $s_\Lambda=1/2$ is the spin of the $\Lambda$ particle.
We assume that the initial and the final states are dominated by configurations
with the $\Lambda$ particle in the $p$ orbits, and we take the orbital
angular momentum of the $\Lambda$ particle to be $l_\Lambda=1$.

The $E2$ transition operator, $\hat{T}_2$,
acts only on the core states. Using
Eq. (7.1.7) in Ref. \cite{Edmonds57},
the reduced
matrix element of the $E2$
operator between the initial and the final states reads,
\begin{eqnarray}
\langle J_f||\hat{T}_2||J_i\rangle
&=&
(-1)^{L_f+\frac{1}{2}+J_i+2}\,\hat{J_i}\hat{J_f}
\left\{\begin{array}{ccc}
L_f & J_f & 1/2 \\
J_i & L_2 & 2
\end{array}\right\} \nonumber \\
&& \times (-1)^{I_f+l_\Lambda+L_i+2}\,\hat{L_i}\hat{L_f}
\left\{\begin{array}{ccc}
I_f & L_f & l_\Lambda \\
L_i & I_i & 2
\end{array}\right\} \nonumber \\
&&\times \langle I_f||\hat{T}_2||I_i\rangle.
\end{eqnarray}
Here, we have used a shorthanded notation of $\hat{J}\equiv \sqrt{2J+1}$.
The $B(E2)$ value for the transition from
the initial to the final states is then given as
\begin{eqnarray}
B(E2; J_i\to J_f) &=&\frac{1}{2J_i+1}\,
|\langle J_f||\hat{T}_2||J_i\rangle|^2, \\
&=&(2J_f+1)(2L_i+1)(2L_f+1) \nonumber \\
&& \times \left\{\begin{array}{ccc}
L_f & J_f & 1/2 \\
J_i & L_i & 2
\end{array}\right\}^2
\left\{\begin{array}{ccc}
I_f & L_f & l_\Lambda \\
L_i & I_i & 2
\end{array}\right\}^2 \nonumber \\
&&\times |\langle I_f||\hat{T}_2||I_i\rangle|^2.
\end{eqnarray}
Notice that
$|\langle I_f||\hat{T}_2||I_i\rangle|^2$ in the last line is related to
the $B(E2)$ value for the core transition as
$|\langle I_f||\hat{T}_2||I_i\rangle|^2 = (2I_i+1)\,B(E2; I_i\to I_f)$.
In order to evaluate it, we use
the collective model~\cite{Ring80}, that is,
\begin{equation}
B(E2; I+2 \to I) = Q_0^2\,\frac{5}{16\pi}\,\frac{3}{2}\,
\frac{(I+1)(I+2)}{(2I+3)(2I+5)},
\end{equation}
for $K=0$, where $Q_0$ is the intrinsic quadrupole moment of a deformed
nucleus.
One then obtains,
\begin{eqnarray}
B(E2; J_i\to J_f) &\propto&
(2J_f+1)(2L_i+1)(2I+1) \nonumber \\
&& \times \left\{\begin{array}{ccc}
L_f & J_f & 1/2 \\
J_i & L_i & 2
\end{array}\right\}^2
\left\{\begin{array}{ccc}
I & L_f & l_\Lambda \\
L_i & I+2 & 2
\end{array}\right\}^2 \nonumber \\
&&\times \frac{(I+1)(I+2)}{(2I+3)},
\end{eqnarray}
for $I_i=I+2$ and $I_f=I$.

Table~\ref{table:B(E2)} summarizes the $B(E2)$ values.
Here, the $B(E2)$ values are given relative to the one
for the transition from the first $5/2^-$ to the first $1/2^-$ states,
which is $7/9$ times $B(E2; 2^+ \to 0^+)$
in the core nucleus.
Only the values larger than 0.25 are listed. The transition strengths
are also graphically shown in Fig. \ref{fig:B(E2)}.
The formation of the band structures can be clearly seen in the figure.
The inter-band transitions are much stronger than
the intra-band transitions, except for the low-lying states.
That is, as expected, the $E2$ cascade transitions are exclusively strong
within the stretched angular momentum states with spin-up (or spin-down),
particularly because the $E2$ operator is spin-independent.

It should be pointed out that the $B(E2)$ values for the states shown in
Fig. \ref{fig:B(E2)} are derived based on the simple picture that only one rotational band is taken into account for the core nuclei.
In our actual calculations for the $_\Lambda$Sm isotopes,
three states ($n_{\rm max}=3$) for a given angular momentum $I$ are adopted, which is much more realistic.


%

\end{document}